\documentclass[aps,prl,reprint,nofootinbib,superscriptaddress,floatfix]{revtex4-2}

\usepackage{color}
\usepackage[T1]{fontenc} 
\usepackage{lmodern}
\usepackage{url}
\usepackage{graphicx}
\usepackage[caption=false]{subfig}
\usepackage{amsmath,amssymb}
\usepackage{amsfonts}
\usepackage{rotating}
\usepackage{epstopdf}
\usepackage{epsfig}
\usepackage{latexsym}
\usepackage{bm}
\usepackage[utf8]{inputenc}
\usepackage{float}
\usepackage{ulem}
\usepackage{hyperref}
\hypersetup{ setpagesize=false,  colorlinks=true, linkcolor=blue, citecolor=red, linktocpage=true, urlcolor=blue, hypertexnames=true}
\usepackage{natbib}

\newcommand{\dphiv}{\dot{\phi}_{0}}
\newcommand{\dchiv}{\dot{\chi}_{0}}

\def\<{\langle\,}
\def\>{\,\rangle}

\newcommand{\Fig}[1]{figure~\ref{#1}}
\newcommand{\Eq}[1]{Eq.~(\ref{#1})}
\newcommand{\Eqs}[2]{eqs.~(\ref{#1}) and (\ref{#2})}

\graphicspath{{figures/}}

\begin{document}

\title{
Primordial Clocks within Stochastic Gravitational Wave Anisotropies
}	
\author{Arushi Bodas}
\email[]{arushib@terpmail.umd.edu}
\author{Raman Sundrum}
\email[]{raman@umd.edu}
\affiliation{Maryland Center for Fundamental Physics, University of Maryland, College Park, MD 20742}


\begin{abstract}
A first order phase transition in the early universe can give an observable stochastic gravitational background (SGWB), which will necessarily have primordial anisotropies across the sky. 
In multi-field inflationary scenarios, these anisotropies may have a significant isocurvature component very different from adiabatic fluctuations, providing an alternate discovery channel for high energy physics at inflationary scales.
Here, we consider classically oscillating heavy fields during inflation that can imprint distinctive scale-invariance-breaking features in the power spectrum of primordial anisotropies. 
While such features are highly constrained in the cosmic microwave background, we show that their amplitude can be observably large in isocurvature SGWB, despite both probing a similar period of inflation.
Measuring SGWB multipoles at the required level, $\ell \sim {\cal O}(10-100)$, will be technologically challenging.
However, we expect that early detection of a strong isotropic SGWB, and the guarantee of anisotropies, 
would spur development of next generation detectors with sufficient sensitivity, angular resolution, and foreground discrimination.
\end{abstract}

\maketitle


\par Detection of gravitational waves (GW) by LIGO/Virgo from  mergers of black-holes and neutron stars \cite{LIGOScientific:2016fpe, LIGOScientific:2017vwq}  has ushered in an era of GW astronomy. 
Apart from localised sources, future GW experiments could also detect diffuse stochastic gravitational wave backgrounds (SGWB).
There are various well-motivated sources of SGWB in the early universe, such as inflationary fluctuations \cite{Starobinsky:1979ty}, particle production \cite{Cook:2011hg}, (p)reheating \cite{Khlebnikov:1997di}, cosmic strings \cite{Vachaspati:1984gt}, and first order phase transitions (PT) \cite{Steinhardt:1981ct, Kosowsky:1992rz, Kamionkowski:1993fg, Nicolis:2003tg, Grojean:2006bp, caprini2016science, Schwaller:2015tja}, many of which can be of detectable strength. While the universe becomes transparent to light only after recombination, the universe is ``transparent'' to GW throughout its history. 
In this way, GW detectors may play a critical role in the observational  cosmology of the very early universe. 

In particular, many extensions of the Standard Model (SM) motivated by the Electroweak Hierarchy Problem would have PTs with critical temperatures $T_{c}\sim (1-1000)$ TeV. 
The GW spectrum from such a PT would then be redshifted to frequencies $\sim$ mHz - Hz today, which fortuitously is the target frequency range of  proposed GW experiments such as LISA \cite{amaro2017laser}, MAGIS \cite{MAGIS-100:2021etm}, TianQin \cite{TianQin:2015yph}, Taiji \cite{Ruan:2018tsw}, BBO \cite{Harry:2006fi}, DECIGO \cite{Kawamura:2011zz}. 
The
GW frequency spectrum is sensitive to the {\it subhorizon} particle physics properties of the PT plasma. 
There will necessarily also be anisotropies $> 10^{-5}$ in the SGWB  which will reflect inflationary {\it superhorizon} fluctuations \cite{Geller:2018mwu} analogous to those of the cosmic microwave background (CMB) (see \cite{baumann2009tasi} for a review of inflation and CMB). 

There is however a striking physical difference between CMB and PT-SGWB anisotropies when it comes to isolating the primordial inflationary fluctuations, due to the very different times of production.
In the CMB,  multipole $\ell_{\rm CMB} \sim 30$ divides subhorizon modes that have re-entered the horizon by the time of recombination, $\ell > \ell_{\rm CMB}$, and superhorizon modes that have not, $\ell < \ell_{\rm CMB}$. 
Subhorizon modes are strongly modified by causal physics
such as baryon acoustic oscillations and Silk damping, and potentially any unknown non-standard physics, while the  superhorizon modes suffer from  significant cosmic variance $\propto 1/\sqrt{\ell}$. 
But because a SGWB is produced much earlier than the CMB, the corresponding dividing line is at a much larger multipole, $\ell_{\rm GW} \sim (T_{c}/T_{\rm CMB}) \, \ell_{\rm CMB} \sim 10^{14}$ for a TeV-scale PT. Thus, 
the large-$\ell$ SGWB anisotropies can simultaneously   have small cosmic variance and be relatively unmodified by subhorizon physics. This, in principle, makes them a pristine and precise map of primordial fluctuations.

On the other hand, measuring such a high-precision SGWB map faces significant practical challenges in terms of detector sensitivity, angular resolution, and subtraction of astrophysical foregrounds. 
Various strategies have been proposed to make progress on these fronts \cite{Cornish:2001hg, Cutler:2005qq, Corbin:2005ny, Kudoh:2005as, Contaldi:2020rht, Biscoveanu:2020gds, Banagiri:2021ovv, baker2021high, Renzini:2022alw}.
If the monopole moment of any SGWB is discovered in relatively near-term detectors such as LISA, there will be strong motivation to build next-generation detectors capable of extracting the {\it guaranteed} anisotropies. 
In this way, we foresee a multi-stage discovery of SGWB anisotropies, analogous to the history of CMB exploration.

We can then ask what new physics might be revealed within an SGWB map. 
Ref.~\cite{Geller:2018mwu} pointed out that the SGWB anisotropy map could have a significant isocurvature component, unlike the  adiabatic perturbations seen in the CMB and large scale structure (LSS),  which is plausible within multi-field inflation and reheating scenarios. Ref.~\cite{Kumar:2021ffi}  demonstrated that such an isocurvature component of the SGWB could even contain significantly Non-Gaussian (NG) correlations, reflecting sizeable interactions among the isocurvature  fields during inflation. 
Another exciting possibility is the detection of very heavy particles/fields, with masses of inflationary energy scales, potentially as high as $10^{15}$ GeV, generally known as ``cosmological collider physics'' \cite{Chen:2009zp, Chen:2012ge, Baumann:2011nk, Noumi:2012vr, Pi:2012gf, Assassi:2012zq, arkani2015cosmological}. In this paper, we study the simplest such phenomenon in the context of SGWB maps. 
 
If the classical inflationary background has (approximate) de Sitter geometry,
\begin{align}
    ds^2 = -dt^2 + a^{2}(t) \, d\vec{x}^2 \, ; \quad a(t) = e^{H t} \,,
\end{align}
its isometries include ``scale invariance'', $\vec{x} \rightarrow \lambda \vec{x}, Ht \rightarrow Ht - \ln \lambda$, where $H$ is the inflationary Hubble scale. 
This relates heavy particle propagation during inflation to spatial correlations at the end of inflation, $e^{-i M (t - t')} \rightarrow (k'/k)^{-i (M/H)}$, where $\vec{k}$ and $\vec{k}'$ are co-moving wave-vectors. 
Such a striking non-analytic signature, sensitive to a heavy mass $M$ can only be seen within NG (greater than two-point) correlators, so that there are at least two independent wave-vectors given the spatial translation invariance.

However, the requirement of NG can be evaded if the classical background contains a small breaking of scale invariance,
which can occur if the heavy field is excited classically, instead of quantum mechanically as above.
Sudden excitation of the heavy field at time $t_{0}$
breaks scale invariance, and the subsequent decay of this field into light quanta allows these scale-invariance-breaking features
to be imprinted in two-point correlators, with a single co-moving wave-vector \cite{Chen:2011zf}. 
The  heavy oscillations then give
$\sim e^{-i M (t - t_{0})} \rightarrow (k_{0}/k)^{-i (M/H)}$, where $k_{0}$ is related to the co-moving mode that is exiting the horizon at $t_{0}$. 
Furthermore, while quantum production of heavy particles in NG is typically ``Boltzmann''-suppressed $\sim e^{-\pi (M/H)}$ (with some special mechanisms that can mitigate this to a limited extent \cite{Chen:2018xck, wang2020gauge, Bodas:2020yho, tong2022large, Kumar:2019ebj, Lu:2019tjj, Chua:2018dqh, Tong:2018tqf}), this is not the case for heavy classical excitations. 
Such an oscillating heavy field has been dubbed a ``primordial clock'' (PC) in the context of testing inflation and alternate scenarios \cite{Chen:2014joa}. Here, we will restrict to the inflationary paradigm (with $\approx$ de Sitter geometry).

Such PCs have not been detected in the precision measurements of CMB by Planck \cite{Akrami:2018odb, Braglia:2021rej},
constraining the couplings of heavy fields to the inflaton (or the field sourcing adiabatic perturbations), and almost saturating the theoretical lower limit from cosmic variance in the low-multipole regime. 
Couplings to isocurvature fields, however, could be significantly larger, giving large PC features in the isocurvature component of SGWB anisotropies.
A SGWB map thus provides a possibility of an alternate channel for detecting heavy field dynamics during inflation.
We construct a model illustrating this idea.

\section{PC in single-field inflation}
We first review the effects of a PC field $\sigma$ on adiabatic density perturbations seeded by the quantum fluctuations $\delta \phi$ of the inflaton field, as proposed in \cite{Chen:2011zf}.  
The initial condition for the classical motion is that it is restricted to slow-roll in the $\phi$ direction of field space, while the massive $\sigma$ field is stuck at its origin. It is assumed that there is a feature in the scalar potential that leads to a sudden transfer of energy to excite $\sigma$ at time $t_0$, after which $\phi$, however, continues to slow-roll.
The classical dynamics of $\sigma$ after $t_0$ is governed by the homogeneous equation of motion,
\begin{align}
    \partial^{2}_{t}\sigma + 3H\, \partial_{t} \sigma+ M^2 \sigma =0 \, ,
\end{align}
where $M$ is the mass of the $\sigma$ field. 
In the limit $M \gg H$, the solution has an oscillatory form,
\begin{align}\label{eq:sig_oscil}
    \sigma_{\rm osc}(t)  \simeq  \theta(t-t_0) \, \sigma_{0}
     \left(\frac{a(t)}{a(t_0)}\right)^{-3/2} \cos[M (t-t_0)  ] \,.
\end{align}
The oscillating field energy in an inflationary background imparts a small oscillatory correction to the Hubble constant,
\begin{align}\label{eq:Hosc}
        \Delta H_{\rm osc} \approx \theta(t-t_0) \frac{\sigma_{0}^2 M}{8 M_{\rm Pl}^2} \left(\frac{a(t)}{a(t_0)}\right)^{-3} \cos[2M (t-t_0)] \, ,
\end{align}
where $M_{\rm Pl}$ is the Planck scale.    
This in turn imparts an oscillatory component to the slow-roll parameters, $\Delta \epsilon_{\rm osc} = -\partial_{t} (\Delta H_{\rm osc})/H^{2}$ and $\Delta \eta_{\rm osc} \approx \partial_{t} (\Delta  \epsilon_{\rm osc})/(H \epsilon)$. Here $H$ and $\epsilon$ are the unperturbed parameters in the absence of $\sigma$ oscillations.

These oscillatory components then introduce scale-invariance-breaking features in the power spectrum of adiabatic perturbations, ${\cal P}_{\zeta_{\phi}}$, defined as
\begin{align}
    \< \zeta_{\phi}(\vec{k}) \zeta_{\phi}(\vec{k}') \> \equiv (2\pi)^3 \delta(\vec{k}+\vec{k}') {\cal P}_{\zeta_{\phi}}(k) \, ,
\end{align}
where $\zeta_{\phi}$ are the gauge-invariant perturbations capturing $\delta \phi$, as measured in the primordial fluctuations of the CMB and LSS. 
This \textit{irreducible} contribution is predominantly given by
\begin{align}\label{eq:cmbSIBgrav}
    \left(\frac{\Delta {\cal P}_{\zeta_{\phi}}}{{\cal P}^{(0)}_{\zeta_{\phi}}}\right)_{\Delta H_{\rm osc}}  \simeq & \,\, \theta(k-k_g) \frac{\sqrt{\pi}}{2} \frac{( M^{2} \sigma_{0}^{2})}{ \,\dphiv^2} \left(\frac{M}{H}\right)^{1/2} \nonumber \\
    & \quad \times \left( \frac{k}{k_g}\right)^{-3} \cos\left[ \frac{2M}{H} \log \frac{k}{k_g}\right] \,.
\end{align}
where ${\cal P}^{(0)} $ denotes the unperturbed scale-invariant spectrum $\propto 1/k^3$.
We can understand the key features of this  calculation intuitively, by noting that ordinarily ${\cal P}_{\zeta_{\phi}} \propto 1/\epsilon$, so that we expect roughly
$\Delta {\cal P}_{\zeta_{\phi}}/{\cal P}^{(0)}_{\zeta_{\phi}} \propto \Delta \epsilon_{\rm osc}/\epsilon$. 
Now, $\Delta \epsilon_{\rm osc}(t)$ represents a high-frequency ($2M$ by Eq.~(\ref{eq:Hosc})) gravitational source for production of inflaton pairs, so their physical momentum on production, $k e^{-Ht} \approx M$, by the local Minkowski energy conservation.
The softest co-moving momentum $k_g$ produced from this gravitational source is given by $k_g  e^{-H t_0} \approx M$. Solving for $t, t_0$ in terms of $k, k_g$ results in the dependence on the latter seen in Eq.~(\ref{eq:cmbSIBgrav}).

Similarly, $\Delta \eta_{\rm osc} $ gives the dominant irreducible PC contribution to the 3-point correlators, which can be written in a following normalized form: 
\begin{align}
 F  \equiv \frac{5}{6} \frac{\<\zeta_{\phi}(\vec{k}_1) \zeta_{\phi}(\vec{k}_2) \zeta_{\phi}(\vec{k}_3)\>}{{\cal P}_{\zeta_{\phi}}(k_1) {\cal P}_{\zeta_{\phi}}(k_2) + (k_1 \rightarrow k_3) + (k_2 \rightarrow k_3)}\,.
\end{align}
In the equilateral triangle configuration $k_{1} \sim k_2 \sim k_3$,
\begin{align}\label{eq:cmbSIB3pnt}
F_{\Delta H_{\rm osc}} &\simeq  \theta(K-2k_g)  \, f_{\rm NL}^{\rm PC} \nonumber \\
&\qquad \times \left( \frac{K}{2 k_g}\right)^{-3} \cos\left[2 \frac{M}{H} \log \frac{K}{2 k_g}\right] \,,
\end{align}
where $K=\sum_{i=1}^{3} |\vec{k}_i|$, and the amplitude of NG can be evaluated to be $f_{\rm NL}^{\rm PC} \sim \frac{10}{9} \frac{\sqrt{\pi}}{4} \frac{(M^{2} \sigma_{0}^{2})}{\dphiv^2} \left(\frac{M}{H}\right)^{5/2}$.

Apart from these irreducible contributions, the coherent $\sigma$ field may also directly decay to inflaton fluctuations.
Consider the lowest dimensional interaction  consistent with an (approximate) inflaton shift-symmetry, ${\cal L}_{\rm int} \sim (\sigma \, \partial_{\mu} \phi \partial^{\mu} \phi)/\Lambda_{\phi} $. 
The PC contribution to ${\cal P}_{\zeta_{\phi}}$ from the decay of $\sigma_{\rm osc}$ through this interaction can be evaluated using the in-in formalism \cite{Chen:2014cwa, chen2022classical} (see \cite{Weinberg:2005vy} for a review of the in-in formalism),
\begin{align}\label{eq:cmbSIBdec}
   \left( \frac{\Delta {\cal P}_{\zeta_{\phi}}}{{\cal P}^{(0)}_{\zeta_{\phi}}}\right)_{\rm dec} \simeq & \,\, \theta(k - k_d)  \sqrt{2 \pi} \frac{ \sigma_0}{\Lambda_{\phi}}  \left(\frac{M}{H}\right)^{1/2} \nonumber\\ 
   & \quad \times \left( \frac{k}{k_d}\right)^{-3/2} \cos\left[\frac{M}{H} \log \frac{k}{k_d}\right] \,.
\end{align}
Here, $\sigma_{\rm osc}$ directly acts as a high-frequency ($M$ from Eq.~(\ref{eq:sig_oscil})) source for inflaton production instead of the irreducible gravitational source $\Delta H_{\rm osc}$.
Therefore, the $k$-dependence can be obtained using a similar argument given for Eq.~(\ref{eq:cmbSIBgrav}). 
With this source, a co-moving mode $k$ is produced when $k e^{-Ht} \approx M/2$. The softest such mode produced through direct decay is then $k_d e^{-Ht_0}=M/2 $. 
Since the interaction is linear in $\sigma$, 
the contribution in Eq.~(\ref{eq:cmbSIBdec}) matter-dilutes 
 $\propto k^{-3/2}$, as compared to the $\propto k^{-3}$ dependence in \Eq{eq:cmbSIBgrav}.



Such PCs have been searched for in the Planck CMB data (see e.g. analysis in \cite{Chen:2014cwa,Braglia:2021rej}).
While there are potential candidates, none are statistically significant. 
In the low to medium $k$-range, the constraints on the amplitudes of PC features in adiabatic perturbations are close to the ultimate cosmic variance limit. 
While future probes like CMB polarization, LSS, and 21-cm tomography will improve sensitivity, they are typically  effective for PCs at high $k$ (small scales)  \cite{Braglia:2021rej, Chen:2016vvw, Chen:2016zuu}. 
The PCs could be undetectable if $\sigma_0$ is  small, or, for the direct decay contribution, if $\Lambda_{\phi}$ is  large.  
Considering the valuable information encoded in the PC features, 
we ask if they can be probed by other cosmological observables.\footnote{
Refs.~\cite{Braglia:2020taf, Fumagalli:2020nvq} have also studied PC-like scale-invariance-breaking features in scalar-induced SGWB. 
There are however important qualitative differences: 1) They consider features that appear at very small scales, which can not be probed by CMB as they are highly Silk-damped. In our work, PC occurs on scales at which the CMB is undamped, but are too small there for detection.
2) Their features appear in the {\it frequency} spectrum of SGWB, unlike in the anisotropies in our case. 
} 

\section{PC in multi-field inflation}
In the following, we study one such possibility where the excited $\sigma$ field decays dominantly to another light ``spectator'' field, $\chi$, present during inflation. 
If  $m_{\chi}^{2} \ll H^2$, it will develop an (approximately) scale-invariant spectrum, ${\cal P}^{(0)}_{{\cal S}_{\chi}} \propto 1/k^3$, similar to that of the inflaton.
Here, ${\cal S}_{\chi} \equiv \zeta_{\chi}-\zeta_{\phi}$ is the isocurvature perturbation of $\chi$, independent of the fluctuations of $\zeta_{\phi}$ when the energy density in isocurvature is subdominant.
To naturally maintain the lightness of $\chi$, we again consider a shift-symmetric effective interaction of the form $\mathcal{L}_{\rm int} \sim  (\sigma \, \partial_{\mu}\chi \partial^{\mu} \chi)/\Lambda_{\chi}$, $\Lambda_{\chi} \ll \Lambda_{\phi}$. This gives a PC feature in $P_{{\cal S}_{\chi}}$ completely analogous to Eq.~(\ref{eq:cmbSIBdec}),\footnote{The contribution from $\Delta H_{\rm osc}$ is small, $(\Delta {\cal P}_{{\cal S}_{\chi}}/{\cal P}^{(0)}_{{\cal S}_{\chi}}) \propto (M^2 \sigma_{0}^2)/(H^2 M_{\rm Pl}^{2})$, and hence we only focus on the direct decay of $\sigma$ to $\delta \chi$.}
\begin{align}\label{eq:chiSIB}
   \frac{\Delta {\cal P}_{{\cal S}_{\chi}}}{{\cal P}^{(0)}_{{\cal S}_{\chi}}} &\simeq  \theta(k-k_d)  \alpha
    \left( \frac{k}{k_d}\right)^{-3/2} \cos\left[\frac{M}{H} \log \frac{k}{k_d}\right] \nonumber \\
    \text{where}\quad \alpha &= \sqrt{\frac{2 \pi M}{ H}} \frac{  \sigma_0}{\Lambda_{\chi}}\,.
\end{align}
There are some constraints on  $\Lambda_{\chi}$ to ensure effective field theory (EFT)  control. We require $ (\sigma_0/\Lambda_{\chi}) < 1$ to keep the correction to the kinetic energy of $\chi$ small, and $(\dchiv/\Lambda_{\chi}^{2}),\, M/\Lambda_{\chi} < 1$ to ensure that no energy scales exceed the scale of non-renormalizability at which the EFT must break down.  
The perturbative control of the in-in diagrams requires the PC feature to be small compared to the scale-invariant spectrum, giving a constraint $\alpha <1$.
We see qualitatively that it is possible to have small $\sigma_0$ and large $\Lambda_{\phi}$ so that PC feature is small in ${\cal P}_{\zeta_{\phi}}$, while $\sigma_0/\Lambda_{\chi}$ gives a sizeable PC contribution in ${\cal P}_{{\cal S}_{\chi}}$.

\section{Non-minimal reheating scenario}
We now illustrate how the primordial fluctuations of $\chi$, including sizeable PC, may be probed by a potentially 
observable SGWB.
\begin{figure}[t]
	\centering
		\includegraphics[width=0.9\columnwidth]{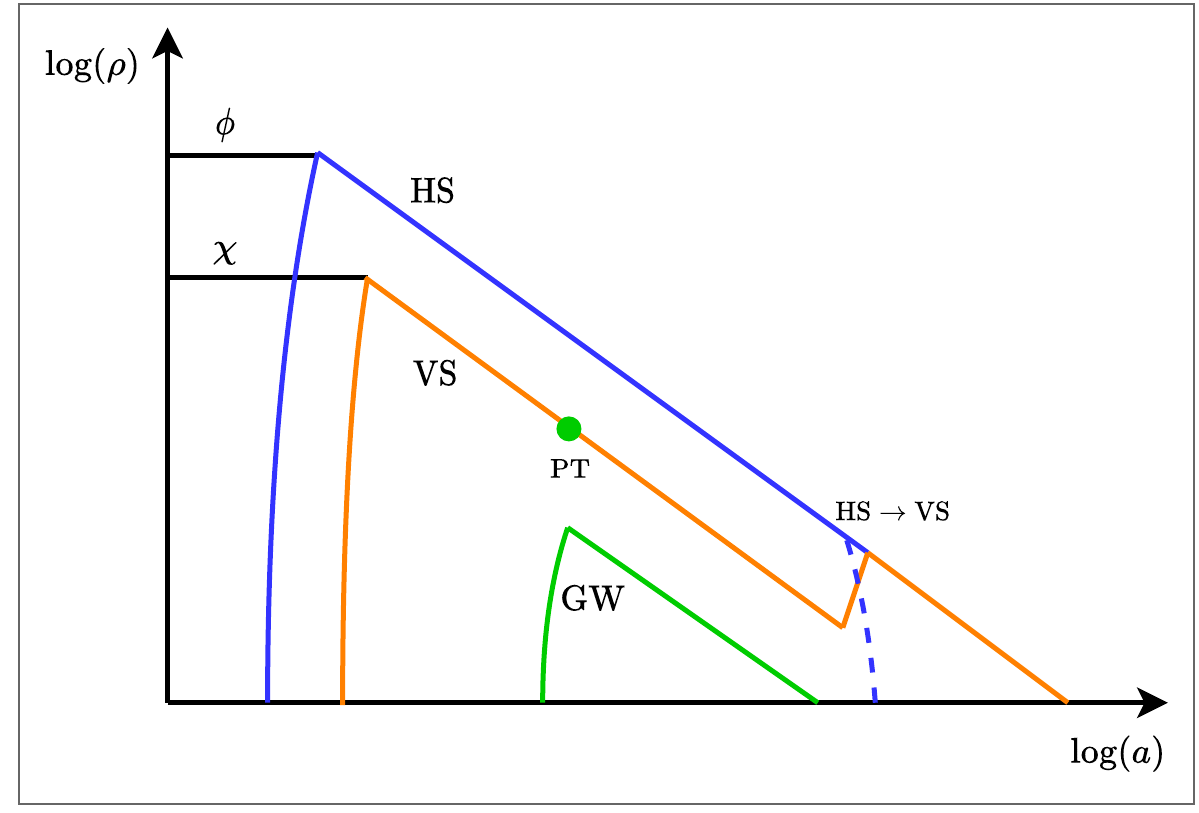}
		\caption{A schematic of the model with modified post-inflationary history showing energy densities in various sectors as a function of the scale factor (time).} 
		\label{fig:ghost}
\end{figure}
We will employ the cosmological model developed in Ref.~\cite{Geller:2018mwu} which showed how  the SGWB can have a large isocurvature component, making it uncorrelated with the CMB. We will add the new features of the $\sigma$ oscillations, back-reaction on the inflationary expansion and decays into $\chi$, perturbatively.
The model is shown schematically in \Fig{fig:ghost}. 
After the end of inflation, $\chi$ reheats the Visible Sector (VS), which is presumed to undergo a strong first order PT, and whose remnants are  just the SM and dark matter (DM). If the PT occurs as the universe is passing through temperatures of order (multi-)TeV, this would result in a SGWB at frequencies observable at proposed GW detectors. 
The inflaton on the other hand is taken to reheat a Hidden Sector of particle physics (HS) that interacts very weakly with the VS.
Some time after the PT, the HS particles decay to the VS, thermalizing with it before  Big Bang Nucleosynthesis. This leaves the SGWB as the only source of isocurvature. 
The constraints from isocurvature and $N_{\rm eff}$ can be cast in terms of the ratio of energy density to critical density today as $\delta \Omega_{\rm GW} h^2 \lesssim 2\times 10^{-11}$ and  $\Omega_{\rm GW} h^2 \lesssim 2\times 10^{-6} $, respectively.
We will later show that our benchmarks easily satisfy these constraints.

The GW are sourced by  sub-horizon physics at the time of PT, and hence the fluctuations $\delta \rho_{\rm GW}$ on super-horizon scales are directly related to the initial fluctuations
in the VS after it is reheated by $\chi$ (but before the HS decays into it), 
\begin{align}
    \delta_{\rm GW} \sim (\delta_{\rm VS})_{\rm initial} \sim \frac{ \delta \chi}{\chi_{0}} \,,
\end{align}
where $\delta \equiv \delta \rho / \rho$ and $\chi_{0}$ is the original homogeneous VEV of the $\chi$ field.
More precisely, the GW anisotropy can be written as \cite{Kumar:2021ffi, Malik:2008im}
\begin{align}\label{eq:delGW}
    \delta_{\rm GW} = -\frac{4}{3}\zeta_{\phi}+4 {\cal S}_{\chi} \left(1-\frac{4}{3}f_{\rm VS}\right) \approx 4 {\cal S}_{\chi} \,,
\end{align}
where we consider $f_{\rm VS} \ll 1$ and  ${\cal S}_{\chi} \gtrsim \zeta_{\phi}$.

The CMB anisotropies receive two independent contributions since the final fluctuations in the VS are a combination of the decay products of the HS and the initial VS,
\begin{align}
    \delta_{\gamma} \sim \left( \delta_{\rm VS} \right)_{\rm final} \sim  \left(\delta_{\rm HS}\right)_{\rm initial} + f_{\rm VS} \left(\delta_{\rm VS}\right)_{\rm initial} \,.
\end{align}
Here, $f_{\rm VS} \equiv (\rho_{\rm VS})_{\rm initial}/\rho_{\rm total}$. 
In terms of the independent gauge-invariant fluctuations,
\begin{align}\label{eq:ghostCMBpert}
    \delta_{\gamma} \approx -\frac{4}{5} \zeta_{\phi} -\frac{4}{5}f_{\rm VS} {\cal S}_{\chi}.
\end{align}
From Eq.~(\ref{eq:ghostCMBpert}), we see that $f_{\rm VS} \ll 1$ can ensure a subdominant contribution of ${\cal S}_{\chi}$ to CMB anisotropies. The GW map in this case will be (mostly) uncorrelated with the CMB since the ${\cal S}_{\chi}$ is independent of $\zeta_{\phi}$. 
An important subtlety of the multi-field reheating is that 
the early Sachs-Wolfe (SW) effect suppresses CMB anisotropies but not the GW anisotropies, $(\delta_{\rm GW}/\delta_{\gamma}) \approx 5 ({\cal S}_{\chi}/\zeta_{\phi})$, as can be seen from Eqs.~(\ref{eq:delGW}) and (\ref{eq:ghostCMBpert}).
This is because the SW effect is {\it anti-correlated} with the dominant adiabatic density fluctuations from $\phi$, leading to reduced anisotropy in the observed CMB \cite{sachs1967perturbations}. On the other hand, the SW contribution, which is predominantly sourced by $\rho_{\rm HS}$, is {\it uncorrelated} with $\chi$ fluctuations, and hence there is no analogous cancellation in GW anisotropies.


Considering the high precision of CMB measurements, we must consider the PC feature induced by ${\cal S}_{\chi}$ into the CMB after the HS decay into the VS, Eq.~(\ref{eq:ghostCMBpert}), as well as
the irreducible PC feature,  Eqs.~(\ref{eq:cmbSIBgrav}) and (\ref{eq:cmbSIB3pnt}), despite their subdominance.
From Eq.~(\ref{eq:ghostCMBpert}), ${\cal S}_{\chi}$ induces
\begin{align}\label{eq:cmbSIBind}
    \left(\frac{\Delta {\cal P}_{\gamma}}{{\cal P}_{\gamma}} \right)_{\rm ind} \approx f^{2}_{\rm VS} \left(\frac{A_{\chi}}{A_{\phi}} \right) \left(\frac{\Delta {\cal P}_{{\cal S}_{\chi}}}{{\cal P}_{{\cal S}_{\chi}}} \right) \,,
\end{align}
where
${\cal P}^{(0)}_{j}(k) = 2 \pi^2 A_j k^{-3}$.
The amplitude $A_{\phi} \approx 2 \times 10^{-9}$ is fixed by the CMB.
An analogous relation holds for PC in LSS.

The anisotropies for a 2D map (CMB or PT-SGWB) are expressed in the multipole basis,
\begin{align}
    \langle \delta(\hat{n}) \delta(\hat{n}') \rangle
    =  \sum_{\ell} \frac{(2 \ell +1)}{4 \pi} C_{\ell} P_{\ell}(\cos \theta)\, , 
\end{align}
where  $\hat{n}\cdot\hat{n}' = \cos \theta$ and $P_{\ell}$ are the Legendre polynomials.
For higher multipoles $\ell \gtrsim O(100)$ (which will be of interest to us later), the expression for $C_{\ell}$ can be simplified using flat-sky approximation (see for example \cite{Munoz:2015eqa}): 
\begin{align}\label{eq:flatskyCl}
    C_{\ell} \approx \frac{1}{r^{2}} \int \frac{d k_{\parallel}}{2\pi} {\cal P}_{\delta}\left(k_{\parallel}, k_{\perp} \sim \frac{\ell}{r} \right) \,,
\end{align}
where ${\cal P}_{\delta}$ is the power spectrum of the density perturbations, $k_{\parallel (\perp)}$ is the component of $\vec{k}$ parallel (perpendicular) to the line-of-sight, and  $r$ is the co-moving radius from us to the surface of the map that we see today. 
The correlation can be separated into a nearly scale-invariant part and a small PC correction, $ C_{\ell} = C^{(0) }_{\ell} + \Delta C_{\ell}$, where $C^{(0)}_{\ell} [\ell (\ell+1)] \equiv A$ is approximately constant. From eqs.~(\ref{eq:delGW}) and (\ref{eq:flatskyCl}), $A_{\rm GW} \simeq 32 \pi A_{\chi}$. 
The PC feature in SGWB can be computed using \Eqs{eq:chiSIB}{eq:flatskyCl}, 
\begin{align}\label{eq:chiSIBl}
    \frac{\Delta C^{\rm GW}_{\ell}}{C^{(0) {\rm GW}}_{\ell}} \approx & \,\, \theta(\ell-\ell_d)\,  \tilde{\alpha}  \left(\frac{\ell}{\ell_d}\right)^{-3/2} \cos\left[ \frac{M}{H} \log \frac{\ell}{\ell_d}\right]  \,,
\end{align}
where the amplitude of the PC feature in the multipole basis, $\tilde{\alpha} = \sqrt{2\pi H/M} \, \alpha$, is smaller than $\alpha$ due to the 2-dimensional nature the map. 
$\ell_d \approx k_d r_{\rm GW}$ corresponds to the multipole where the PC feature appears in the GW map.
We can similarly evaluate the amplitude of PCs in CMB using eqs.~(\ref{eq:cmbSIBgrav}), (\ref{eq:cmbSIBdec}), and (\ref{eq:cmbSIBind}) instead.


With sufficient experimental sensitivity, the detection of the PC feature is ultimately limited by  cosmic variance.
If we search in the parameter regime $H \ell_d/M \gg 1$, then our signal oscillates slowly as a function of $\ell$, in which case we can mitigate cosmic variance by averaging the data over a bin of size $1 < \Delta \ell < \pi H \ell_d/M$ around each $\ell$. 
Thus the signal adds coherently while the binned cosmic variance is
\begin{align}\label{eq:cvbin}
    \frac{\delta C_{\ell}^{bin}}{C_{\ell}} \approx  \sqrt{\frac{2}{(2\ell+1) \Delta \ell}} \,.
\end{align}
We determine the observability of the PC feature by comparing its largest amplitude (at $\ell_d$) with the smallest compatible  binned cosmic variance,
\begin{align}\label{eq:alphaObs}
     \tilde{\alpha} > \sqrt{\frac{2 (M/H)}{\pi (2\ell_d +1) \ell_d}} \, .
\end{align}
PCs with smaller amplitudes can be observable if they appears at larger $\ell_d$, as the cosmic variance drops. It is however experimentally challenging to probe high $\ell$ anisotropies. Keeping these issues in mind, we will consider $\ell_d \sim 100$ in this work. 

We also want to estimate the strength of GW signal in our model.
During PT, processes like bubble collisions, sounds waves, and turbulence give rise to GW \cite{Caprini:2015zlo}. 
After production, GW decouple and redshift as radiation until today.
An estimate for the peak energy density today as a fraction of critical energy density \cite{Hindmarsh:2013xza} is \footnote{We have only considered bubble wall collisions here, but the contribution from sound waves is generally comparable and even dominates at larger $\beta$ as it scales linearly with ($H_{\rm PT}/\beta$).}
\begin{align}\label{eq:gwMonopole}
    \Omega_{\rm GW} h^2 \sim 1.3 \times 10^{-6} \left(\frac{H_{\rm PT}}{\beta}\right)^2 f_{\rm VS}^{2} \,.
\end{align}
Here $\beta$ is the rate of bubble nucleation, $H_{\rm PT}$ is the Hubble expansion rate during PT, and we are considering a strong PT where most of the latent heat goes into accelerating the bubble walls. 
The factor of $f_{\rm VS}$ appears since the PT is occurring in the VS sector that is subdominant in the energy density during PT.

The energy density fluctuation in the $\ell^{th}$ multipole is given by
\begin{align}\label{eq:gwAnisotropy}
    \delta \Omega_{\rm GW} h^2 (\ell) &\sim \Omega_{\rm GW} h^2 \times \frac{\sqrt{A_{\rm GW}}}{\ell}
\end{align}
The factor of $1/\ell$ appears because $C^{(0)}_{\ell} \propto [\ell (\ell +1)]^{-1}$ by scale-invariance. 
The quantity of interest to us is the deviation from this scale-invariant spectrum due to the PC feature, $\Delta C_{\ell}^{\rm GW}$. 
From Eq.~(\ref{eq:chiSIBl}),
\begin{align}\label{eq:gwSIB}
     \delta \Omega_{\rm GW}^{\rm PC} h^2 (\ell_d) \sim
    & 1.3 \times 10^{-6} \left(\frac{H_{\rm PT}}{\beta}\right)^2 f_{\rm VS}^{2} \frac{\sqrt{A_{\rm GW}}}{\ell_d} \left(\frac{\tilde{\alpha}}{2}\right).
\end{align}

\section{Benchmarks}
\begin{figure*}[ht]
    \subfloat[\footnotesize{Unbinned}]{\includegraphics[width=0.45\linewidth]{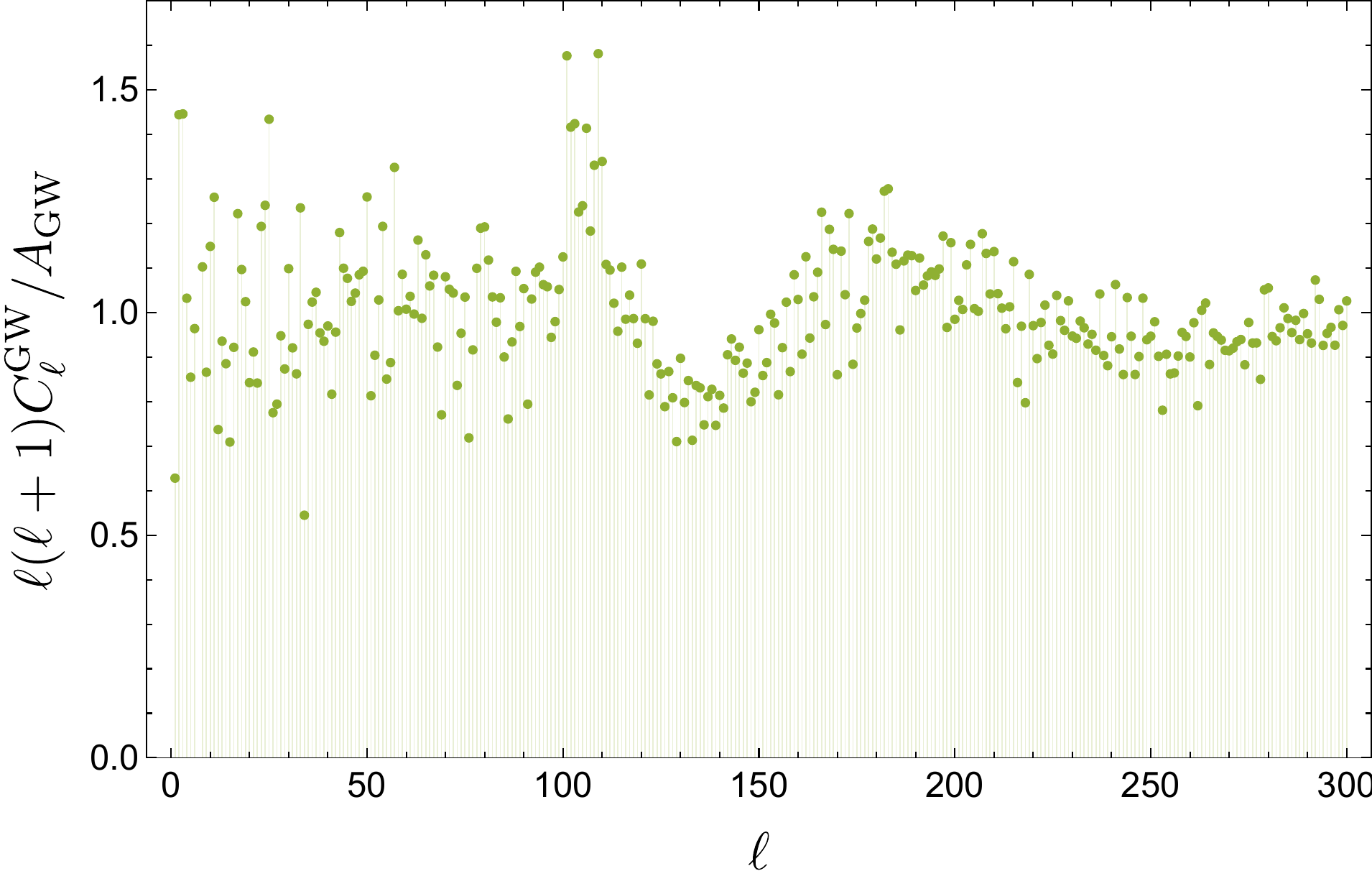}}
    \hspace{3em}
    \subfloat[\footnotesize{Binned, $\Delta \ell = 6$}]{\includegraphics[width=0.45\linewidth]{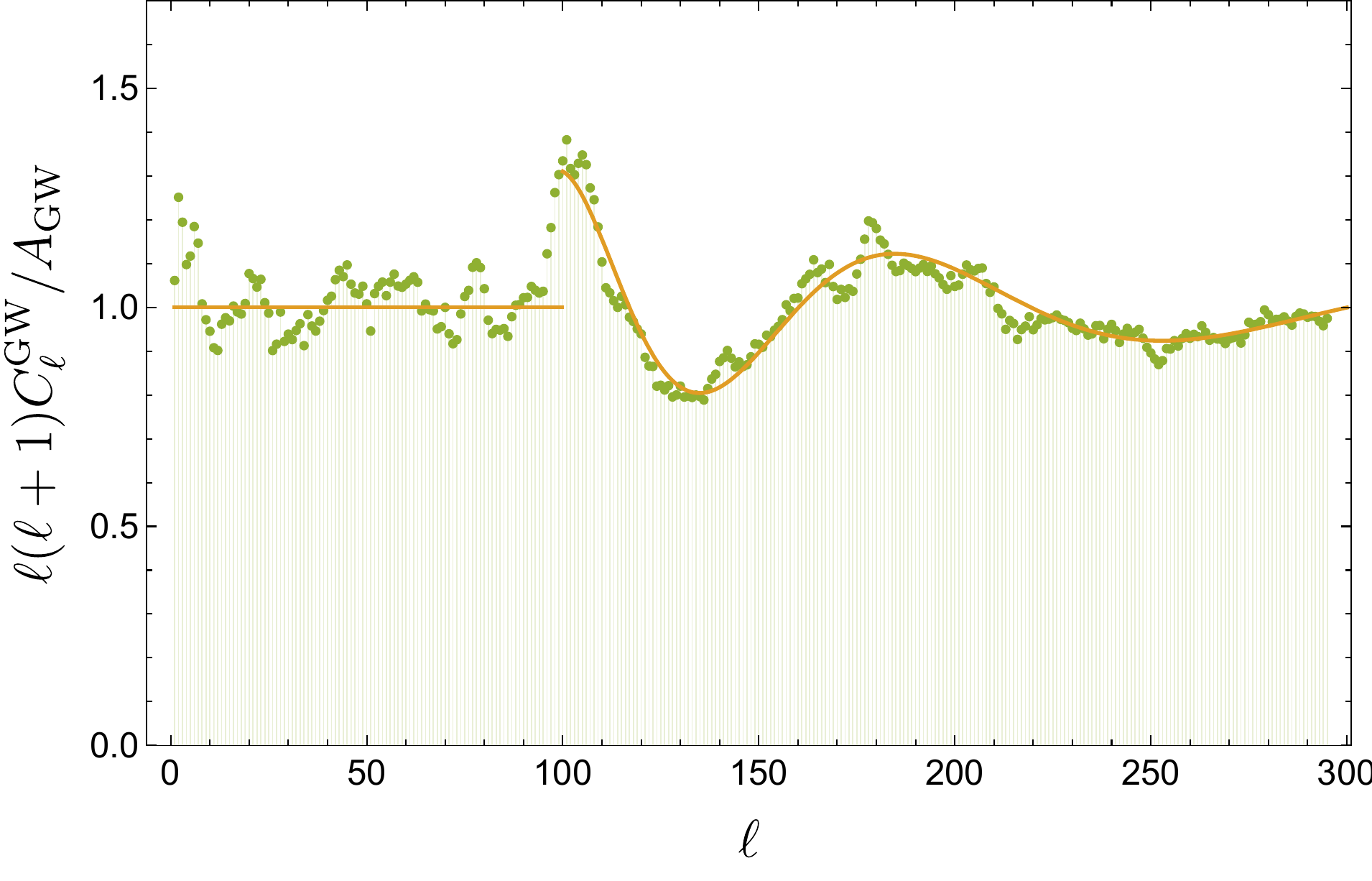}}
	\caption{Primordial spectrum $C^{\rm GW}_{\ell}$ with the PC feature using benchmark parameters from Eq.~(\ref{eq:benchmark-common}). The yellow line is the expected signal where the onset of PC feature is happening at $\ell_d=100$ and the green dots incorporate a realization of cosmic variance. Figures (a) and (b) correspond to GW map without and with binning ($\Delta \ell = 6$) respectively.
    } 
	\label{fig:benchmark_plots}
\end{figure*}

We will now illustrate our mechanism with two benchmarks.
Consider the following set of parameters,
\begin{align}\label{eq:benchmark-common}
    \sigma_0 \simeq 2H ,\quad \Lambda_{\chi} \simeq 40 H,\quad   M = 10H , \quad  \ell_d = 100 
\end{align}
Let us first consider the PC signal in SGWB.
From eqs.~(\ref{eq:chiSIB}) and (\ref{eq:chiSIBl}), the amplitude of the PC feature is $\tilde{\alpha} = 0.31$.
This easily satisfies the condition of observability over cosmic variance, $\tilde{\alpha} > 0.018$, given in Eq.~(\ref{eq:alphaObs}).
A realization of SGWB map containing this PC feature and a random sample of cosmic variance is shown in Fig.~(\ref{fig:benchmark_plots}) (a) without binning and (b) with binning using $\Delta \ell = 6$. We see that even with a modest bin size, the observability of signal is significantly improved, with the signal-to-noise ratio $\sim 7.6$.

There are also irreducible PC corrections to the maps of adiabatic perturbations as discussed before.
Currently, the strongest constraints on these corrections come from the CMB \cite{Braglia:2021rej}. 
Therefore, in our first benchmark, we consider a PC whose contributions in the 2D map of CMB are completely hidden under  cosmic variance, and yet are  observable in the SGWB.
In the future, 3D maps such as LSS and 21-cm are expected to achieve higher sensitivity as they can access more modes than a 2D map, and hence have lower cosmic variance.
It is then interesting to ask whether there are PCs which can not be detected even in a perfect 3D map of adiabatic perturbations, making isocurvature SGWB possibly the only channel of discovery.
We demonstrate this possibility in the second benchmark.

Consider the first benchmark along with Eq.~(\ref{eq:benchmark-common}),
\begin{align}
    (1) \quad f_{\rm VS} = 0.1, \, \frac{A_{\chi}}{A_{\phi}}  = 3 , \, \frac{\beta}{H_{\rm PT}} = 5 \,.
\end{align}
The PC corrections in CMB also appear around $\ell_{d}$ and $\ell_g$, as the co-moving radii to the surface of recombination and PT are approximately the same. 
We take $\Lambda_{\phi} \gg H$ such that the contribution in Eq.~(\ref{eq:cmbSIBdec}) is negligible.
Eq.~(\ref{eq:cmbSIBgrav}) along with $\dphiv \simeq (60H)^2$ \cite{Akrami:2018odb} gives a correction $\left(\Delta C^{\gamma}_{\ell_g} / C^{(0) \gamma}_{\ell_g}\right)_{\Delta H_{\rm osc}} \simeq 6 \times 10^{-5}$, and
Eq.~(\ref{eq:cmbSIBind}) gives $\left(\Delta C_{\ell_d}^{\gamma} / C_{\ell_d}^{(0)\gamma}\right)_{\rm ind} \simeq 0.01$.
This is below the cosmic variance as can be checked with Eq.~(\ref{eq:alphaObs}) (even after accounting for a smaller cosmic variance at $\ell_{g} \approx 200$).
This benchmark is actually conservative in that the current constraints on the PC features from Planck data are even weaker \cite{Braglia:2021rej}, only constraining the amplitudes of PC features to be $\lesssim 0.08$.
Access to higher multipoles $\ell \sim {\cal O} (1000)$ in the CMB is not really an advantage as the $\ell^{-3/2}$ (or $\ell^{-3}$) dependence in the PC features makes them fairly localised near $\ell_d$ (or $\ell_g$).

We now estimate the experimental sensitivity required to probe this PC in SGWB.
The strength of SGWB depends on the details of the PT.
Here, we consider a relatively strong PT
(see \cite{Caprini:2015zlo} for models that could undergo strong PTs).
From Eqs.~(\ref{eq:gwMonopole}) and (\ref{eq:gwSIB}), we get the isotropic component, $\Omega_{\rm GW} h^2 \simeq 5  \times 10^{-10} $, and the small PC variation in anisotropies, $\delta \Omega_{\rm GW}^{\rm PC} h^2 (\ell_d) \simeq 4 \times  10^{-15}$. 
Note that the isocurvature and $N_{\rm eff}$ bounds mentioned earlier are easily satisfied.
While the isotropic SGWB could be within LISA sensitivity, $\Omega_{\rm GW}^{\text{\tiny LISA}} h^2 \gtrsim 10^{-14}$ \cite{caprini2016science}, 
the anisotropies require more advanced future space-based detectors.
For reference, the projected sensitivity of the proposed BBO mission for large angular scales is $\Omega_{\rm GW}^{\text{\tiny BBO}} h^2 \gtrsim 10^{-17}$ \cite{Corbin:2005ny, Schmitz:2020syl} and that of ultimate-DECIGO is $\Omega_{\rm GW}^{\text{\tiny DECI}} h^2 \gtrsim 10^{-19}$ \cite{Kudoh:2005as}.
Detection of the PC will require a detector that maintains such a high sensitivity out to $\ell \sim O(100)$.

Now consider a second benchmark with smaller $f_{\rm VS}$,
\begin{align}
       (2) \quad f_{\rm VS} = 0.01, \, \frac{A_{\chi}}{A_{\phi}}  = 3 , \, \frac{\beta}{H_{\rm PT}} = 5 \,.
\end{align}
The PC corrections in this case from eqs.~(\ref{eq:cmbSIBgrav}) and (\ref{eq:cmbSIBind}) are respectively $\left(\Delta {\cal P}_{k_g} / {\cal P}_{k_g}^{(0)}\right)_{\Delta H_{\rm osc}} \simeq 8 \times 10^{-5}$ and
$\left(\Delta {\cal P}_{k_d} / {\cal P}_{k_d}^{(0)}\right)_{\rm ind} \simeq 10^{-4}$.
3D maps of adiabatic perturbations like LSS and 21-cm are expected to have lower cosmic variance as they can access more modes than the 2D CMB.
Cosmic variance for an idealised 3D map is $\left(\delta {\cal P}/{\cal P}^{(0)}\right) \sim \sqrt{(2 k_{\rm min}^{3})/(4 \pi k^{2} \Delta k)}$, where $k_{\rm min}$ in the largest co-moving mode that can be probes today.
Then, using the largest possible bin size, $\Delta k \sim \pi k H/M$, the cosmic variance at $k_d \approx 100 k_{\rm min}$ is $\left(\delta {\cal P}/{\cal P}^{(0)}\right) \sim 7\times 10^{-4}$, and at $k_g \approx 200 k_{\rm min}$ is $\left(\delta {\cal P}/{\cal P}^{(0)}\right) \sim 2\times 10^{-4}$.
We see that the PC correction in the second case can not be detected even with an ideal 3D map. 

The strength of SGWB in this benchmark is smaller, with monopole, $\Omega_{\rm GW} h^2 \simeq 5  \times 10^{-12} $, and the PC feature, $\delta \Omega_{\rm GW}^{\rm PC} h^2 (\ell_d) \simeq 4 \times  10^{-17}$.
While this case is more challenging experimentally, the impossibility of detection even in a perfect adiabatic map makes SGWB a powerful (and possibly the only) probe of such a PC.

The final check is the NG contribution to adiabatic perturbations as shown in \Eq{eq:cmbSIB3pnt}. In both benchmarks, the imparted NG is $f_{\rm NL}^{\rm PC} \sim 5 \times 10^{-3}$, which is well below the CMB constraint $f_{\rm NL} < {\cal O}(5-50)$ (the exact constraint depends on the shape) \cite{Akrami:2019izv}. 
A conservative estimate of the cosmic variance for a 3D map gives $\delta f_{\rm NL}^{\rm PC} \sim 1/(\sqrt{N_{k_g}} \zeta_{\phi}) \sim 10^{-2}$, where $N_{k_g}$ is the number of 3-point configurations satisfying $k_1+k_2+k_3 = 2 k_g$ and $\vec{k_1}+\vec{k_2}+\vec{k_3}=0$, and the anisotropy $\zeta_{\phi} \sim 10^{-5}$. 
Again, the imparted NG in our benchmarks is safely below  cosmic variance.

\begin{figure*}[ht]
    \subfloat[\footnotesize{$\ell_d=100$}]{\includegraphics[width=0.45\linewidth]{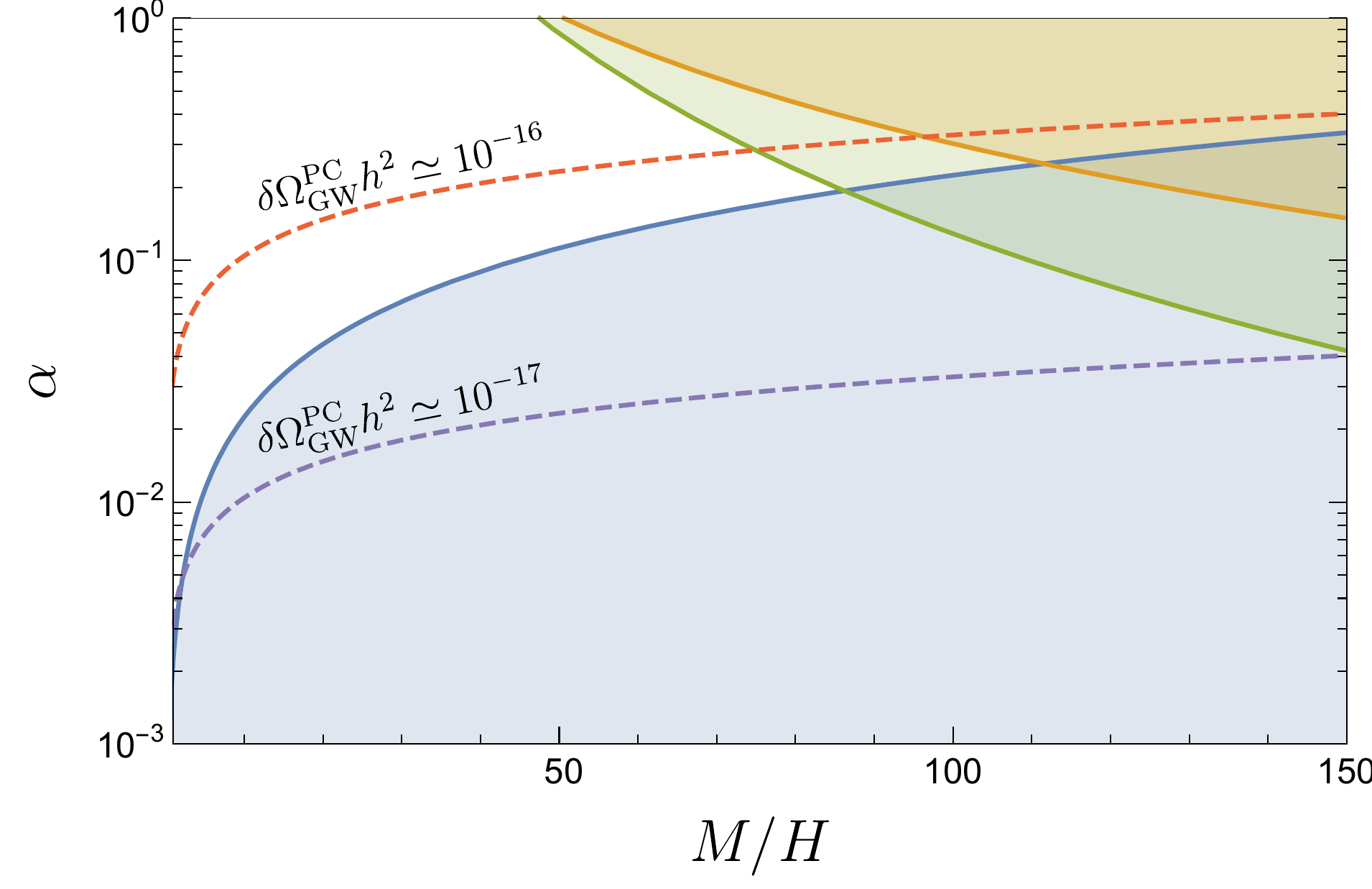}}
   \hspace{3em}
    \subfloat[\footnotesize{$\ell_d=1000$}]{\includegraphics[width=0.45\linewidth]{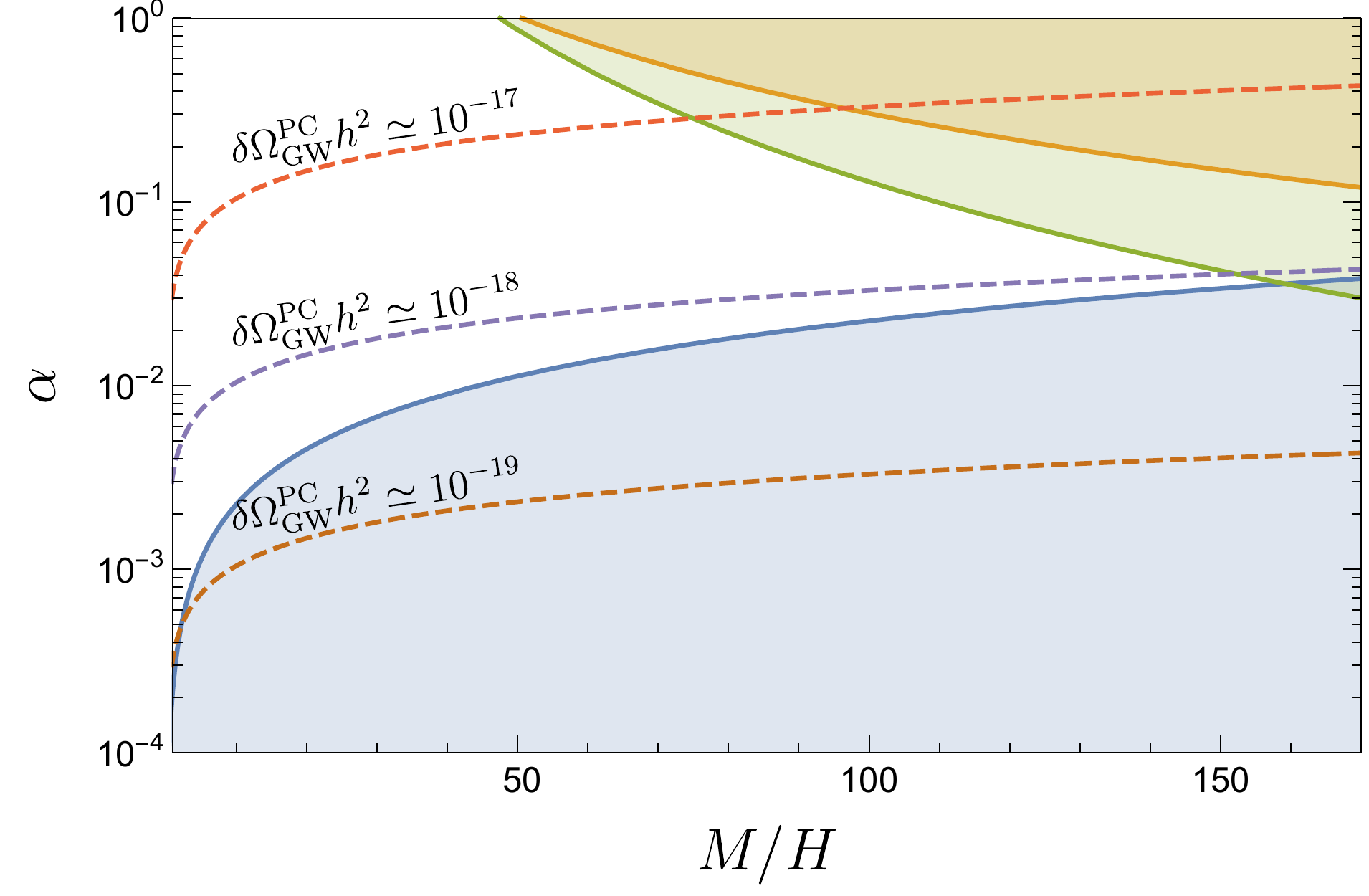}}
	\caption{
	Analysis of the $\alpha-M$ parameter space showing excluded regions from following considerations: binned cosmic variance (blue), CMB constraints from contribution to $\Delta {\cal P}_{\gamma}$ (yellow) and $f_{\rm NL}$ (green).
	The dashed lines show contours of PC signal strength $\delta \Omega_{\rm GW}^{PC} h^2 (\ell_d)$ as given in Eq.~(\ref{eq:gwSIB}) 
	such that the parameter regions above these lines can be probed with corresponding experimental sensitivities. 
	The analysis is carried out for (a) $\ell_d =100$, as in the benchmark cases, and (b) $\ell_d =1000$, where the cosmic variance uncertainty is smaller.
	See the text for more details. 
    } 
	\label{fig:parameter_plot}
\end{figure*}

\section{broader analysis of the parameter space}
Let us now broaden  the exploration of the  parameter space of PC in SGWBs beyond the specific benchmarks above.
We continue to fix the start of the PC feature at $\ell_d=100$, as in the benchmark cases, and analyze observability of PC in $\alpha-M$ space, taking both of these as our independent variables. 
The requirement of $\alpha <1$ and $M>H$ ensures $(\sigma_0/\Lambda_{\chi}) < 1$, satisfying the EFT perturbativity constraint.

Requiring that the (envelope) of the PC feature is larger than the binned cosmic variance  excludes the blue region shown in figure~(\ref{fig:parameter_plot}). 
Requiring Eqs.~(\ref{eq:cmbSIBgrav}) and (\ref{eq:cmbSIB3pnt}) to lie below current CMB sensitivity, $(\Delta {\cal P}/{\cal P}^{(0)})_{\gamma} < 0.01$ and $f_{\rm NL} < 10$, along with  
the restriction 
$M < \Lambda_{\chi}$ needed for EFT control, 
exclude the yellow and green regions respectively in  \Fig{fig:parameter_plot}. 
The contribution from HS decay in Eq.~(\ref{eq:cmbSIBind}) can be kept below CMB exclusion with a conservative choice of $f_{\rm VS}^{2} = 3 \times 10^{-3}$ (for $A_{\chi}/A_{\phi} =3$), while $f_{\rm VS}$ can in principle be larger for smaller $\alpha$. 
The contours of different signal strengths,  $\Omega^{\rm PC}_{\rm GW} h^2$, at given $\ell_d$ are also shown in \Fig{fig:parameter_plot} with dashed lines.

We have assumed that the decay rate $\Gamma(\sigma \rightarrow \delta \chi \, \delta \chi) < 3H$ such that the dilution of the classical $\sigma$ field is primarily due to the Hubble expansion, $\propto k^{-3/2}$ dependence in the PC signal. For the interaction considered, $\Gamma(\sigma \rightarrow \delta \chi \delta \chi) \approx \frac{M^3}{32 \pi \Lambda_{\chi}^{2}}$ \cite{Kofman:1997yn}. 
For the largest mass $M \simeq \Lambda_{\chi}$, we see that $\Gamma < 3H$ for $M\lesssim 300 H$.
Also, the PC oscillation would be unobservable if it is faster than the integer discretization of $\ell$. 
Since the oscillation frequency is fastest near $\ell_d$, we require $(M/H) < \pi \ell_d$.
For $\ell_d = 100$, we again get a similar upper limit $M \lesssim 300H$, although CMB constraints become dominant well before reaching this limit as can be seen in \Fig{fig:parameter_plot}~(a).

At least in principle, PC features with smaller amplitudes can be observable if they appear at larger $\ell_d$, as the cosmic variance becomes smaller at higher multipoles. An example is given in \Fig{fig:parameter_plot}~(b) for $\ell_d=1000$. This highlights the gains that can be achieved with higher angular resolution and sensitivity in GW experiments.

\section{Oscillating heavy field from a polynomial potential}
The above analysis considered an unspecified ``sharp feature'' in the potential \cite{Chen:2011zf,Braglia:2021rej} to excite the heavy field, leading to the simplest form for the PC features. Here we construct a  smoother, more familiar polynomial potential for $\sigma, \phi$, inspired by Hybrid Inflation \cite{Linde:1993cn, Wang:2018tbf}.
In this model, the rolling of the inflaton field triggers a ``waterfall'' for the $\sigma$ field. The waterfall however does not end inflation, but leads to a second phase of inflation.
\begin{align}\label{eq:hybridpot}
    V(\phi, \sigma) = V_{\rm inf}(\phi) - \kappa \sigma+\frac{1}{2}(g\phi^2-\mu^2)\sigma^2+ \frac{\lambda}{4} \sigma^4 \, .
\end{align}
We will take $V_{\rm inf}(\phi)$ to dominate the energy density and hence drive inflation.  $\sigma$ has a time-dependent mass as $\phi$ rolls towards 0,
\begin{align} \label{eq:Msigma}
    M^{2}_{\sigma}(t) = g \phi^{2}(t)-\mu^2 + 3 \lambda \<\sigma\>^2 \,.
\end{align}
Initially, $g \phi^2 \gg \mu^2$ and $\< \sigma \> \approx 0$, $V \approx V_{\rm inf}$. As $\phi$ rolls,  $\sigma$ becomes tachyonic as $g \phi^2 < \mu^2$. The transition occurs at $t_0$ where  $\phi(t_0) = \phi_c$ and
\begin{align}
    g \phi_{c}^{2} = \mu^2 \,.
\end{align}
From here, $\sigma$ starts rolling rapidly towards the new minimum in about $\Delta t \sim H^{-1}$ when the kinetic energy is large enough to overcome Hubble friction. 
When $\sigma$ becomes large enough, the last term of Eq.~(\ref{eq:Msigma}) ensures $\sigma$ is non-tachyonic again and
oscillates around the adiabatic minimum $ M_{\sigma}(t)/\sqrt{2 \lambda}$, which approaches a maximum value $\mu/\sqrt{\lambda}$ at late time. 
Here we are taking $\kappa$ to be small such that its role in the $\sigma$ dynamics after $t_0$ is subdominant. 
Earlier, it breaks the degeneracy of the two $\sigma$ minima when $\sigma$ becomes tachyonic. 
This ensures selection of a unique vacuum ($\<\sigma\> >0$ for $\kappa >0$) in the entire observable universe, avoiding the domain wall problem \cite{Linde:1993cn}\footnote{In \cite{Wang:2018tbf}, this is done by considering $\sigma$ to be the radial direction in the quasi-single field model of inflation. Then the centrifugal force due to the turning trajectory acts as the source of explicit symmetry breaking.}.

Consider the following model parameters: 
$\mu = 50H$, $g=1.9\times10^{-8}$, $\phi_c = 3.6 \times 10^{5} H$, $\lambda = 0.1$, and $\kappa= 0.1 H^3$ (with $H/M_{\rm pl} = 10^{-5}$ and $\dphiv \simeq 3.6 \times 10^{3} H^2$ \cite{Akrami:2018odb}). 
Let us first check that the back-reaction of $\sigma$ dynamics on $\phi$ is small.
The maximum energy density in $\sigma$ is subdominant to the inflationary potential energy, $\mu^4/\lambda \sim 10^{7} H^4 \ll V_{\rm inf} \simeq 3H^2 M_{\rm pl}^{2} \sim 10^{10} H^4$. 
The kinetic energy of $\phi$ is given by $3H \dphiv \approx -\partial_{\phi} V(\phi,\sigma) = -\partial_{\phi}V_{\rm inf} - g \<\sigma\>^2 \phi$. 
The correction $g \<\sigma\>^2 \phi \lesssim O(10) H^3  \ll \partial_{\phi}V_{\rm inf} \sim 10^4 H^3$. This ensures slow roll of the inflaton even after the tachyonic transition in $\sigma$.

\begin{figure}[t]
   \includegraphics[width=0.85\linewidth]{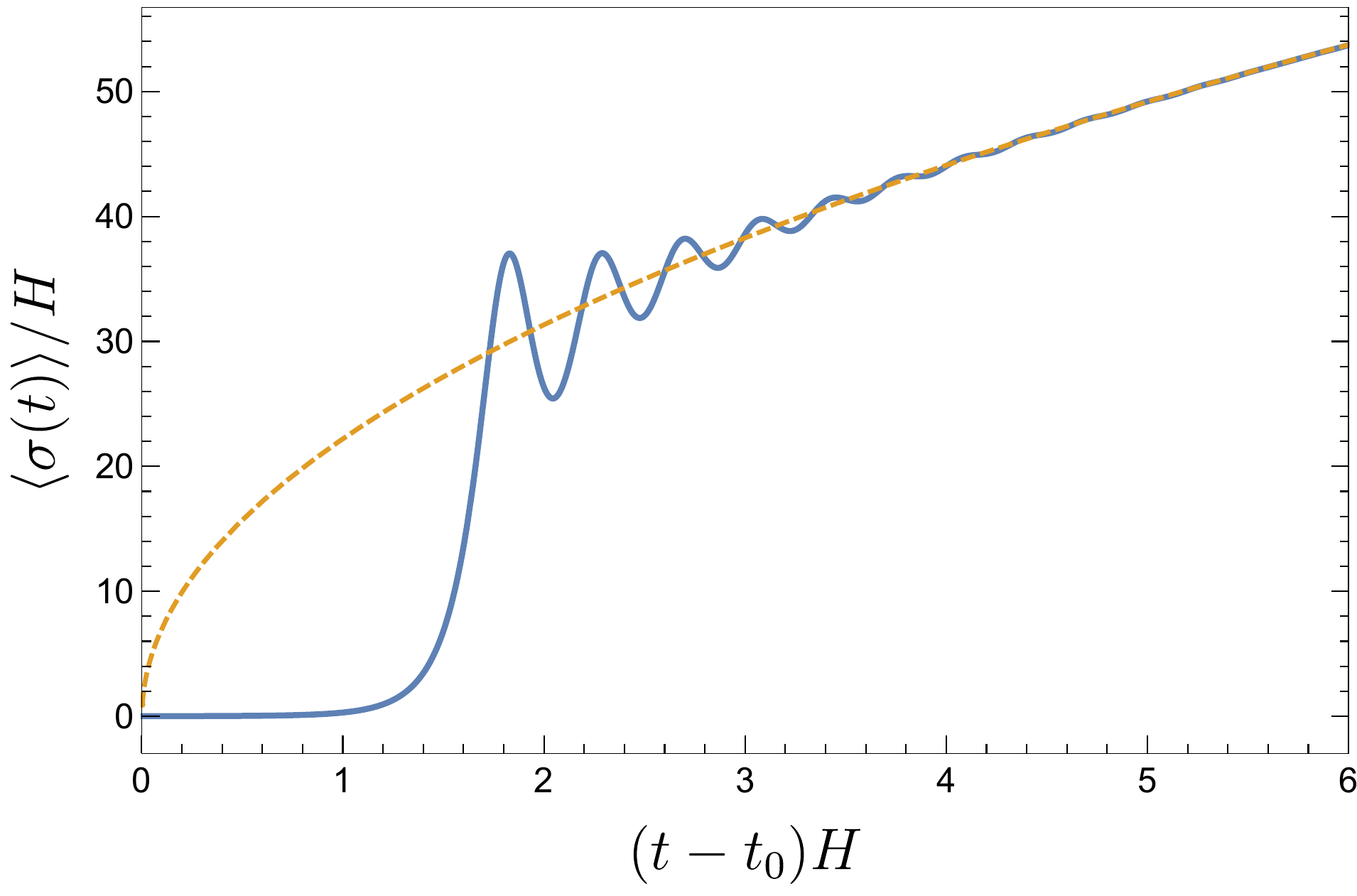}
	\caption{The classical dynamics of $\sigma$ field after the tachyonic transition at $t_0$ (blue) and the minimum $ M_{\sigma}(t)/\sqrt{2\lambda}$ (yellow). The full dynamics shows oscillations about the minimum.} 
	\label{fig:hybrid_sigma}
\end{figure}
\begin{figure}[t]
	\centering
		\includegraphics[width=0.95\linewidth]{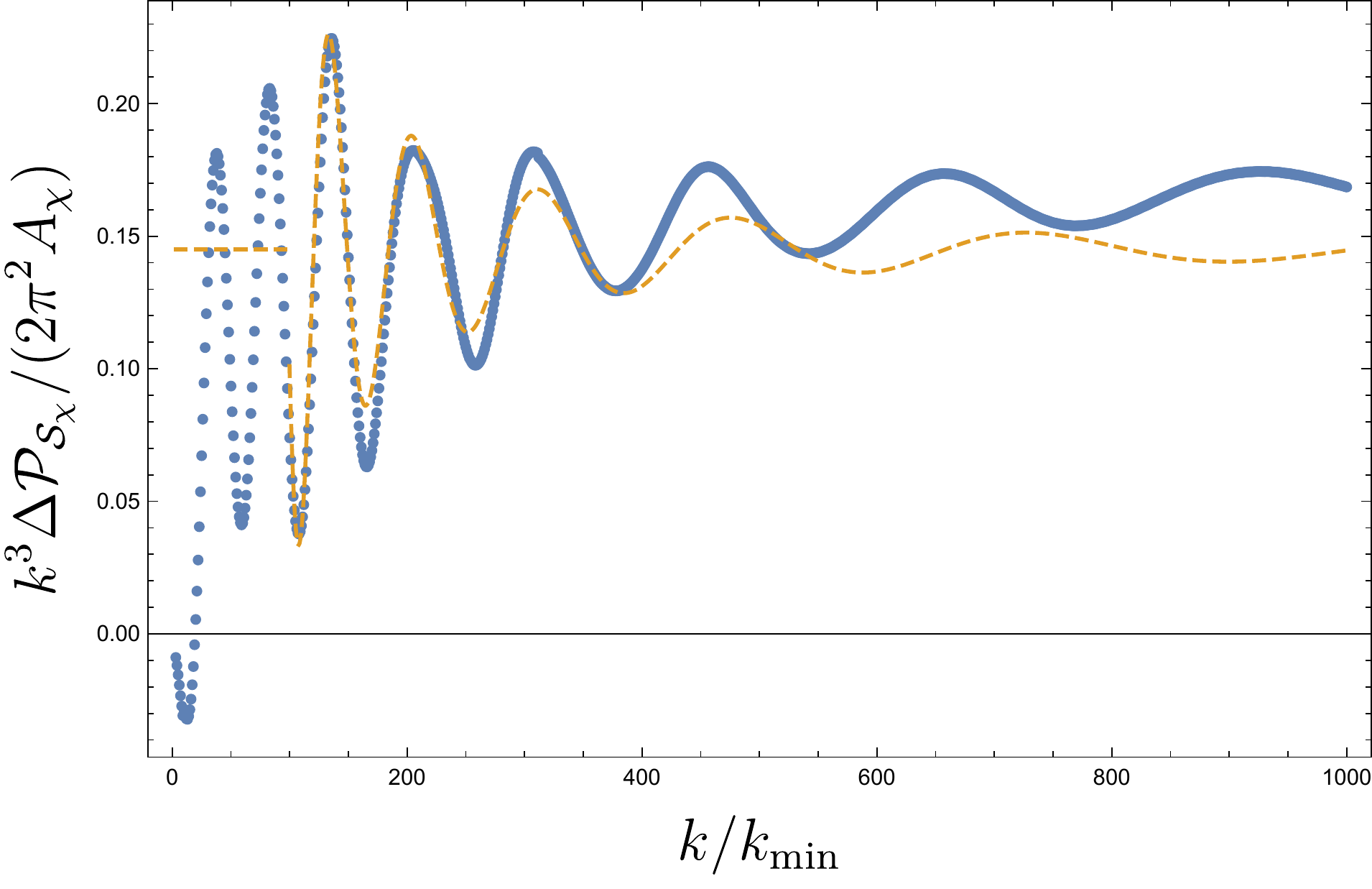}
		\caption{The PC feature in ${\cal P}_{{\cal S}_{\chi}}$ in the model with polynomial potential. The blue dots correspond to the numerical evaluation of $\Delta {\cal P}_{{\cal S}_{\chi}}$, where the overall shift is the result of $\<\sigma\> \neq 0$. For reference, the yellow dashed line shows a matching signal from a simplified scenario where $\sigma$ is instead excited suddenly, resulting in a PC with constant amplitude $\alpha \approx 0.125$ and mass $M \approx 14.8 H$.} 
		\label{fig:hybrid_sig_k}
\end{figure}

In \Fig{fig:hybrid_sigma}, the full $\< \sigma (t)\>$ dynamics is shown in blue along with the expected minimum at $ M_{\sigma}(t)/\sqrt{2\lambda}$ shown in yellow. 
We can see that the dynamics is given by an oscillatory component superimposed on a slowly shifting minimum. 
We consider the same direct coupling to $\chi$ as before, $\mathcal{L}_{\rm int} \sim (\sigma \, \partial_{\mu} \chi \partial^{\mu} \chi)/\Lambda_{\chi}$. The slowly shifting minimum induces a tilt in the otherwise scale-invariant spectrum, while the oscillatory component adds a PC feature.
We take $\Lambda_{\chi} \sim 300 H$ so as to ensure $(\< \sigma \> / \Lambda_{\chi}) < 0.5 $. 

Figure~\ref{fig:hybrid_sig_k} shows the fractional correction to ${\cal P}_{{\cal S}_{\chi}}$ evaluated numerically (blue) and
a matching signal (yellow) of the type considered in the previous sections arising from a more sharply featured potential with $\alpha \approx 0.125$ and $M \approx 14.8 H$.
The overall shift in ${\cal P}_{{\cal S}_{\chi}}$ is the result of $\<\sigma(t)\> \neq 0$.
The largest amplitude of PC feature is around $k_d \sim 100 k_{\rm min}$, corresponding to $\ell_d \sim 100$ as expected. 
The PC is within the observable parameter space, as can be seen from \Fig{fig:parameter_plot}. 
In this model, 
the irreducible PC contributions to the CMB from eqs.~(\ref{eq:cmbSIBgrav}) and (\ref{eq:cmbSIB3pnt}) are $(\Delta {\cal P}/{\cal P}^{(0)})_{\gamma} \simeq 0.001$ and $f_{\rm NL} \simeq 0.1$ respectively, which are below CMB sensitivity.
Comparing to the yellow signal due to sudden excitation of the $\sigma$ field, we see that our waterfall model gives a PC that 
dilutes slightly slower and oscillates with higher frequency as $M_{\sigma}$ increases with time.

\section{Conclusion}
We have shown that the power spectrum of SGWB superhorizon anisotropies could reveal striking oscillations (see figure~(\ref{fig:benchmark_plots})), which could only arise from extremely heavy fields during inflation.
However, any such new physics introduces irreducible (gravitational) corrections to the spectrum of adiabatic perturbations. 
Remarkably,  it is possible for the SGWB oscillations to be clearly visible while the  CMB and LSS remain insensitive to the irreducible corrections, despite their high precision and statistics.
While we have considered specific PC features, the mechanism can be extended to other localised features and to non-Gaussian correlators.
We illustrated a model in which the oscillations in anisotropy appeared at $\ell \sim {\cal O}(100)$ in order to maintain better theoretical control and minimize cosmic variance. Similar features could appear at $\ell \sim {\cal O}(10)$, but would require stronger couplings and a more rigorous data analysis. However, lower experimental angular resolution would then suffice.  In our smoother model, the primordial clock ``beats'' irregularly, but its non-trivial 
frequency $M(k)$ seen in \Fig{fig:hybrid_sig_k}
sheds light on the smooth potential and mechanism exciting $\sigma$.

\textit{Acknowledgements:} 
We would like to thank Soubhik Kumar, Yuhsin Tsai, Naren Manjunath, and Peter Shawhan for useful discussions. This work was supported by NSF grant PHY-1914731 and the Maryland Center for Fundamental Physics.

	\bibliography{refs.bib}
\end{document}